\documentclass[sn-aps,iicol]{sn-jnl}

\usepackage{graphicx}%
\usepackage{multirow}%
\usepackage{amsmath,amssymb,amsfonts}%
\usepackage{amsthm}%
\usepackage{mathrsfs}%
\usepackage[title]{appendix}%
\usepackage{xcolor}%
\usepackage{textcomp}%
\usepackage{manyfoot}%
\usepackage{booktabs}%
\usepackage{algorithm}%
\usepackage{algorithmicx}%
\usepackage{algpseudocode}%
\usepackage{listings}%
\usepackage{lipsum}  
\usepackage{siunitx}  

\usepackage{caption}
\usepackage{subcaption}
\usepackage{amsmath}

\newcommand{\tab}{Table~}
\newcommand{\fig}{Figure~}

\theoremstyle{thmstyleone}%
\theoremstyle{thmstyletwo}%

\theoremstyle{thmstylethree}%

\begin{document}

\title[HERMES payloads performance]{The ground calibration of the \textit{HERMES-Pathfinder} payload flight models}


\author[1]{G.~Dilillo}

\author[2]{E.~J.~Marchesini}

\author[3, 19]{G.~Baroni}

\author[1]{G.~Della~Casa}

\author[2]{R.~Campana}

\author[1, 5]{Y.~Evangelista}

\author[4]{A.~Guzm\'an}

\author[4]{P.~Hedderman}

\author[9]{P.~Bellutti}

\author[16]{G.~Bertuccio}

\author[1]{F.~Ceraudo}

\author[3]{M.~Citossi}

\author[15,17]{D.~Cirrincione}

\author[16]{I.~Dedolli}

\author[9]{E.~Demenev}

\author[1, 5]{M.~Feroci}

\author[9]{F.~Ficorella}

\author[11]{M.~Fiorini}

\author[16,9]{M.~Gandola}

\author[18]{M.~Grassi}

\author[10]{G.~La Rosa}

\author[1]{G.~Lombardi}

\author[18]{P.~Malcovati}

\author[16]{F.~Mele}

\author[10]{P.~Nogara}

\author[1]{A.~Nuti}

\author[6,7]{M.~Perri}

\author[4]{S.~Pliego-Caballero }

\author[8]{S.~Pirrotta}

\author[8]{S.~Puccetti}

\author[13]{I.~Rashevskaya}

\author[10]{F.~Russo}

\author[10]{G.~Sottile}

\author[4]{C. Tenzer}

\author[12,19]{M.~Trenti}

\author[3]{S.~Trevisan}

\author[15,17]{A.~Vacchi}

\author[15]{G.~Zampa}

\author[15]{N.~Zampa}

\author[3]{F.~Fiore}

\affil[1]{\orgname{INAF-IAPS Roma}, \orgaddress{\street{Via del Fosso del Cavaliere 100}, \city{Rome}, \country{Italy}}}
\affil[2]{\orgname{INAF-OAS Bologna}, \orgaddress{\street{Via Piero Gobetti 101}, \city{Bologna}, \country{Italy}}}
\affil[3]{\orgname{INAF-OATS Trieste}, \orgaddress{\street{Via Giambattista Tiepolo 11}, \city{Trieste}, \country{Italy}}}
\affil[4]{\orgname{IAAT-Universit\"at Tübingen}, \orgaddress{\street{ Geschwister-Scholl-Platz}, \city{T\"ubingen}, \country{Germany}}}
\affil[5]{\orgname{INFN Roma Tor Vergata}, \orgaddress{\street{Via della Ricerca Scientifica 1}, \city{Rome}, \country{Italy}}}
\affil[6]{\orgname{INAF-OAS Roma}, \orgaddress{\street{Via Frascati 33}, \city{Monte Porzio Catone}, \country{Italy}}}
\affil[7]{\orgname{ASI Space Science Data Center }, \orgaddress{\street{Via del Politecnico snc}, \city{Rome}, \country{Italy}}}
\affil[8]{\orgname{Agenzia Spaziale Italiana (ASI)}, \orgaddress{\street{Via del Politecnico snc}, \city{Rome}, \country{Italy}}}
\affil[9]{\orgname{Fondazione Bruno Kessler} \orgaddress{\street{Via Sommarive, 18}, \city{Povo}, \country{Italy}}}
\affil[10]{\orgname{INAF-IASF Palermo}, \orgaddress{\street{Via Ugo La Malfa 153}, \city{Palermo}, \country{Italy}}}
\affil[11]{\orgname{INAF-IASF Milano}, \orgaddress{\street{Via Alfonso Corti 12}, \city{Milano}, \country{Italy}}}
\affil[12]{\orgname{University of Melbourne}, \orgaddress{\street{VIC 3010}, \city{Melbourne}, \country{Australia}}}
\affil[13]{\orgname{TIFPA} \orgaddress{\street{Via Sommarive, 18}, \city{Povo}, \country{Italy}}}
\affil[15]{\orgname{INFN Trieste}, \orgaddress{\street{AREA Science Park, Padriciano 99}, \city{Trieste}, \country{Italy}}}
\affil[16]{\orgname{Politecnico di Milano}, \orgaddress{\street{Via Anzani 42}, \city{Como}, \country{Italy}}}
\affil[17]{\orgname{Università degli Studi di Udine}, \orgaddress{\street{via delle Scienze 206}, \city{Udine}, \country{Italy}}}
\affil[18]{\orgname{Università di Pavia}, \orgaddress{\street{Via Ferrata, 5}, \city{Pavia}, \country{Italy}}}
\affil[19]{\orgname{Università degli Studi di Trieste}, \orgaddress{\street{Piazzale Europa, 1}, \city{Trieste}, \country{Italy}}}
\affil[20]{Australian Research Council Centre of Excellence for All-Sky Astrophysics in 3-Dimensions, Australia}

\abstract{
HERMES-Pathfinder is a space-borne mission based on a constellation of six nano-satellites flying in a low-Earth orbit. The 3U CubeSats, to be launched in early 2025, host miniaturized instruments with a hybrid Silicon Drift Detector/scintillator photodetector system, sensitive to both X-rays and gamma-rays. A seventh payload unit is installed onboard SpIRIT, an Australian-Italian nano-satellite developed by a consortium led by the University of Melbourne and launched in December 2023. The project aims at demonstrating the feasibility of Gamma-Ray Burst detection and localization using miniaturized instruments onboard nano-satellites.

The HERMES flight model payloads were exposed to multiple well-known radioactive sources for spectroscopic calibration under controlled laboratory conditions. The analysis of the calibration data allows both to determine the detector parameters, necessary to map instrumental units to accurate energy measurements, and to assess the performance of the instruments. We report on these efforts and quantify features such as spectroscopic resolution and energy thresholds, at different temperatures and for all payloads of the constellation. Finally we review the performance of the HERMES payload as a photon counter, and discuss the strengths and the limitations of the architecture.
}


\maketitle

\section{Introduction}\label{sec:introduction}

The \emph{High Energy Rapid Modular Ensemble of Satellites} (HERMES) is a constellation of nanosatellites to study high-energy astrophysical transients, such as Gamma-Ray Bursts (GRBs)~\cite{fiore2020hermes}.
HERMES will employ the signal triangulation technique to localize high-energy transients. This technique was first utilized by the VELA spacecraft constellation~\cite{klebesadel1973observations}, which first discovered GRBs, and is based on multiple detections by different spacecraft separated by distances comparable to Earth radius. Due to the finite speed of light, detectors hosted by different spacecraft record the transients at slightly different times. The source location is estimated from the measurement of these time delays, with an accuracy which scales with the number of spacecraft and the average baseline length~\cite{sanna2020timing}.

To demonstrate the feasibility of detecting and localizing GRBs using miniaturized instrumentation on nanosatellites, the Italian Space Agency (ASI) funded the HERMES-Pathfinder (\emph{HERMES-Pathfinder} hereafter) mission, which is a constellation made up of six nano-satellites, three funded by ASI (HERMES-TP) and three funded by the European Commission Horizon 2020 Research and Innovation Program (HERMES-SP).  HERMES-Pathfinder constellation is expected to be launched in low-Earth orbit in early 2025~\cite{fiore2020hermes}. 
A seventh payload unit, identical to those hosted by HERMES-Pathfinder, is installed onboard SpIRIT~\cite{auchettl2022spirit}, an Australian-Italian nano-satellite mission developed by a consortium led by the University of Melbourne (Principal Investigator Michele Trenti). 
SpIRIT has been successfully launched into a Sun-synchronous orbit on December 1st, 2023. It is currently undergoing commissioning, with the first gamma and x-ray spectra acquired suggesting a nominal performance.  

The HERMES spacecraft in the HERMES constellation are CubeSats, a class of nanosatellites~\cite{cubesatspecs}. CubeSats adopt a modular design along with standardized sizes and form factors to reduce cost and development time. Each HERMES-Pathfinder spacecraft consists of a three unit (3U) CubeSat, with two units dedicated to the service module and one unit allocated for the scientific payload. The HERMES scientific payload is a miniaturized detector based on the ``siswich" architecture~\cite{fuschino2020innovative}, which enables the realization of a broad-band high-energy detector within a compact form factor by utilizing solid state Silicon Drift Detectors (SDDs). These instruments serve as both an active detector for X-ray radiation and a readout device for the visible light generated when $\gamma$-rays are absorbed by inorganic scintillation crystals. 

Between 2021 and 2023, the payloads of the HERMES-Pathfinder constellation and SpIRIT were assembled, tested, and calibrated. 
For ground calibration, each payload was exposed to a small number of well-known radioactive sources under a variety of operational conditions and in a controlled environment. From the analysis of the data acquired in this process, several instrument parameters were measured. The knowledge of these parameters allows for the accurate energy measurement of photons from arbitrary sources. These efforts and their results are the subject of this work. This paper is organized in three sections. The payload architecture is outlined in Sec.~\ref{sec:payload}, the calibration strategy and methodology are described in Sec.~\ref{sec:calib}, and the results of the calibration are summarized and discussed in Sec.~\ref{sec:results}.

\section{The HERMES payload}\label{sec:payload}

\begin{figure}
\centering
        \begin{subfigure}{0.495\textwidth}
        \centering
        \includegraphics[width=0.7\textwidth]{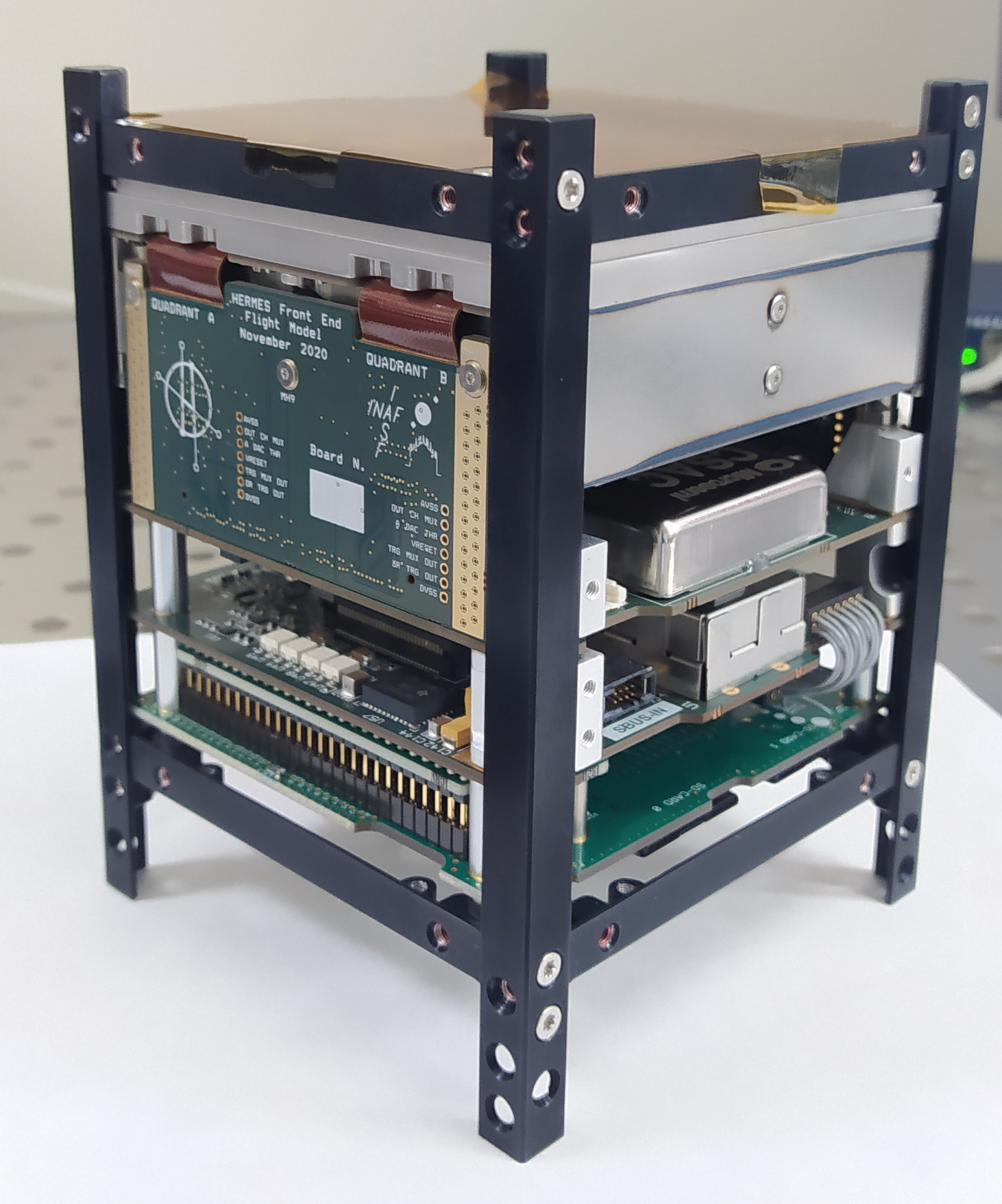}
        \caption{The HERMES-Pathfinder \emph{PFM} payload.}
        \label{fig:plreal}
    \end{subfigure}
    \begin{subfigure}{0.495\textwidth}
        \centering
        \includegraphics[width=0.95\textwidth]{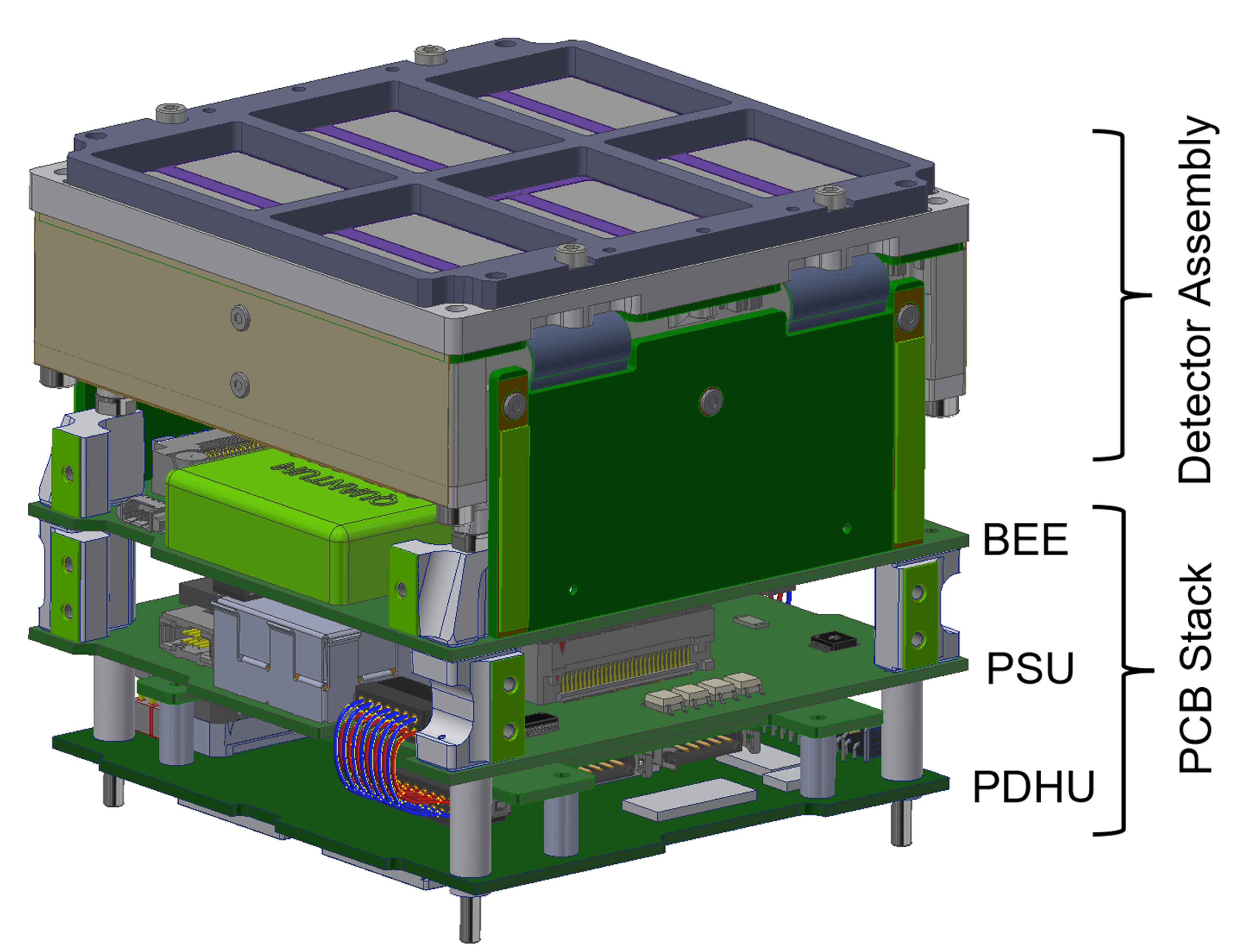}
        \caption{Schematic view of the HERMES payload.}
        \label{fig:plcad}
    \end{subfigure}

\caption{Implementation and design of the HERMES-Pathfinder payload.}
\label{fig:payload}
\end{figure}

The payload of HERMES hosts the detector~\cite{evangelista2022design}, a back-end electronic board (BEE), the payload data handling unit (PDHU, a miniaturized computer for CubeSats)~\cite{guzman2020payload}, and a power-supply board (PSU)~\cite{nogara2022power}, see \fig\ref{fig:payload}. These subsystems are contained within modules of dimensions $10$ cm $\times$ $10$ cm $\times$ $10$ cm, with a mass smaller than $2$ kg, in compliance with the CubeSat standards~\cite{cubesatspecs}. Table \ref{tab:masspowtab} reports the power consumption under scientific observation workload at $20^\circ$C and the mass of each payload unit.

\paragraph{The detector} The HERMES detector is based on the “siswich” architecture~\cite{fuschino2020innovative}.  In a siswich detector, silicon sensors (SDDs) are optically coupled to a scintillator crystal. Soft X-ray photons are absorbed in the silicon bulk and read-out by the detector itself. Hard X-ray and gamma-ray photons have enough energy to pass through the silicon detectors and thus to reach the scintillator, where they are absorbed and a fraction of their energy is converted into optical light. This scintillation light is, in turn, collected by the same silicon detectors.
The discrimination between the two signals is achieved exploiting a segmented design: a single scintillator crystal is coupled to two silicon drift detectors, so that an event detected by a single detector will be associated to a single X-ray photon being absorbed by one of the SDDs (``X" event), while an event detected simultaneously in two adjacent detectors, optically coupled to the same crystal, will be associated to the scintillation light from hard-X and gamma-ray photons (``S" event).
The coincidence window is short enough (nominally $15$~$\mu$s) to ensure a low probability of misclassification, even when observing very bright GRB sources.
The scintillation crystal chosen for the HERMES detector is GAGG:Ce (Gd$_3$Al$_2$Ga$_3$O$_{12}$:Ce, Cerium-doped Gadolinium Aluminium Gallium Garnet). 
GAGG:Ce provides an array of desirable features for spaceborne applications, such as high-radiation tolerance, mechanical robustness and low intrinsic background \cite{dilillo2022space}\cite{della2024new}\cite{yoneyama2018evaluation}. Most importantly, GAGG:Ce has a high density of $6.2$ g/cm$^3$ and high light yield~\cite{kamada20122}. These properties allow GAGG:Ce to effectively stop incident radiation of higher energy compared to lighter materials, ultimately resulting in higher upper-bound thresholds, and to provide high signal-to-noise ratio for low energy photon detection. The HERMES detector utilizes SDDs as the solid-state detector technology~\cite{gatti1984semiconductor}. One main advantage of SDDs is the decoupling of the photon collection area from the region where charges are read out (anode), which can be then minimized in area, allowing to achieve a very low output capacitance value \cite{bertuccio2024anode} and, consequently, very low electronic noise levels. The HERMES payload~\cite{evangelista2020scientific, evangelista2022design} hosts 60 GAGG:Ce scintillation crystals, each with dimensions of $12.10 \times 6.94 \times 15.00$ mm$^3$, coupled to 120 SDDs with a surface area of $7.44 \times 6.05$ mm$^2$ and a thickness of $450$ $\mu$m, distributed across four identical \emph{quadrant} units. The scintillator crystals are housed in a tungsten-shielded box, on top of which a structure for supporting the detectors is placed.

\paragraph{Front-end and back-end electronics, payload data handling unit}

The HERMES  application specific integrated circuit (ASIC) is called LYRA and is an heritage of the VEGA ASIC project developed for the readout of large-area SDDs~\cite{gandola2021multi}\cite{gandola2019lyra}\cite{grassi2020x}\cite{dedolli2023updates}\cite{ahangarianabhari2015vega}\cite{feroci2012large}. The HERMES detectors host four independent LYRA ASICs chipsets, one per quadrant. Each ASIC chipset is divided in two stages, called LYRA-FE (front-end) and LYRA-BE (back-end). The LYRA-FE, whose input is bonded to the anode of a single SDD, is composed by 30 integrated circuits with functions of charge sensitive amplifier, first stage signal shaper and current-mode line driver. 
The nominal shaping time is set to 2~$\mu$s, with a measured jitter less than $100$ ns (1 $\sigma$).
To optimize the noise performance of the SDD-LYRA system, the interconnection and stray capacitances between the detector and the front-end input stage must be kept as low as possible \cite{bertuccio2024electronic}. For this reason, the LYRA-FE integrated circuits are placed in close proximity to the SDDs anodes. We call \emph{channel} the combination of an SDD cell and its FE integrated circuit.
The LYRA-BE is a monolithic multichannel chip and is placed in a farther position with respect to the SDD. It completes the signal processing chain, providing a low-input impedance buffer, the second shaping stage, discriminators, pile-up rejection and peak and hold circuitry. It is also responsible for the BEE-ASIC communication interface and for the multiplexing of the signal from the 30 LYRA-FE input channels stage.

The signals from the four LYRA-BE chips are routed towards the HERMES back-end electronics (BEE). The BEE hosts an FPGA, four analog-to-digital converters (ADC), a miniaturized atomic clock, as well as a number of ancillary components (e.g., pulse generators, temperature sensors). At back-end level the analog signal is digitized and “tagged” with the event amplitude in instrumental units, the detection time and the channel address. 
The timing information is derived from a 100 ns counter, reset at the pulse-per-second (PPS) signal from the on-board GPS receiver for on-board time correction. An ultra-stable chip-scale atomic clock (CSAC) provides $10$~MHz and $1$~Hz clock signals guaranteeing time-stamping accuracy and clock synchronization, even when the GPS is not locked. 
The stability of the atomic clock signal was tested over periods of 3600~s at various temperature levels, ranging from $-$25~$^\circ$C to 20~$^\circ$C. The results showed that the clocks maintained a frequency stability of under 1~ppb \cite{evangelista2022design}, in line with the manufacturer specifications.

The HERMES-TP/SP Payload Data Handling Unit (PDHU) is built around a flight-proven miniaturized computer hosting a 400 MHz ARM9 processor alongside a mass memory and standardized interfaces. A custom-made daughter-board provides all the payload-bus electrical interfaces as well as the PDHU internal interface with the other payload subsystems~\cite{guzman2020payload}. The PDHU runs the software managing the instrument operative modes. It is also responsible for running the online burst search algorithm and preparing the scientific and housekeeping data for telemetry, among other tasks.

\begin{table}
\centering
\begin{tabular}{lrr}
\toprule
 & Mass [g] & Power [W] \\ \midrule
PFM & 1585.51 & 1.685 \\
FM1 & 1562.28 & 1.633 \\
FM2 & 1573.72 & 1.633 \\
FM3 & 1581.09 & 1.695 \\
FM4 & 1575.63 & 1.675 \\
FM5 & 1592.37 & 1.677 \\
FM6 & 1579.10 & 1.680 \\
\bottomrule
\end{tabular}
\bigskip
\caption{Payloads mass and power consumption under scientific operation workload at +20~$^\circ$C. Measure errors are $10$ mg and $1$ mW. FM1 - installed on SpIRIT - has a slightly smaller mass compared to other payloads due to the differences between the mechanical and thermal interfaces of the SpIRIT and the HERMES platforms.} 
\label{tab:masspowtab}
\end{table}

\section{Calibration campaign}
\label{sec:calib}

\begin{figure*}
\centering
\includegraphics[width=\textwidth]{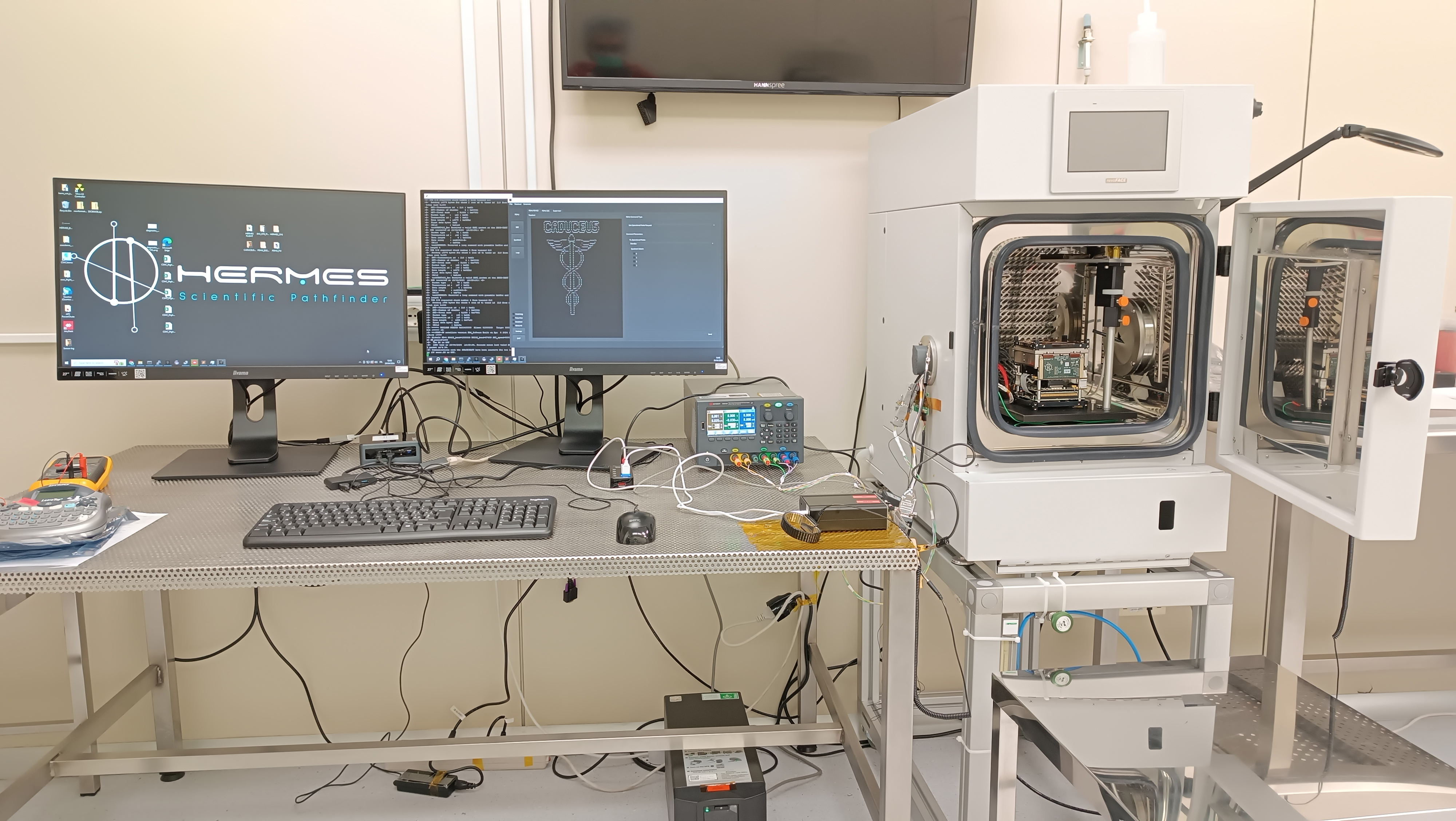}
\caption{The calibration experimental setup. The payload is in the climatic chamber, on the right.}
\label{fig:real_setup}
\end{figure*}

\begin{figure}
\centering
\includegraphics[width=\columnwidth]{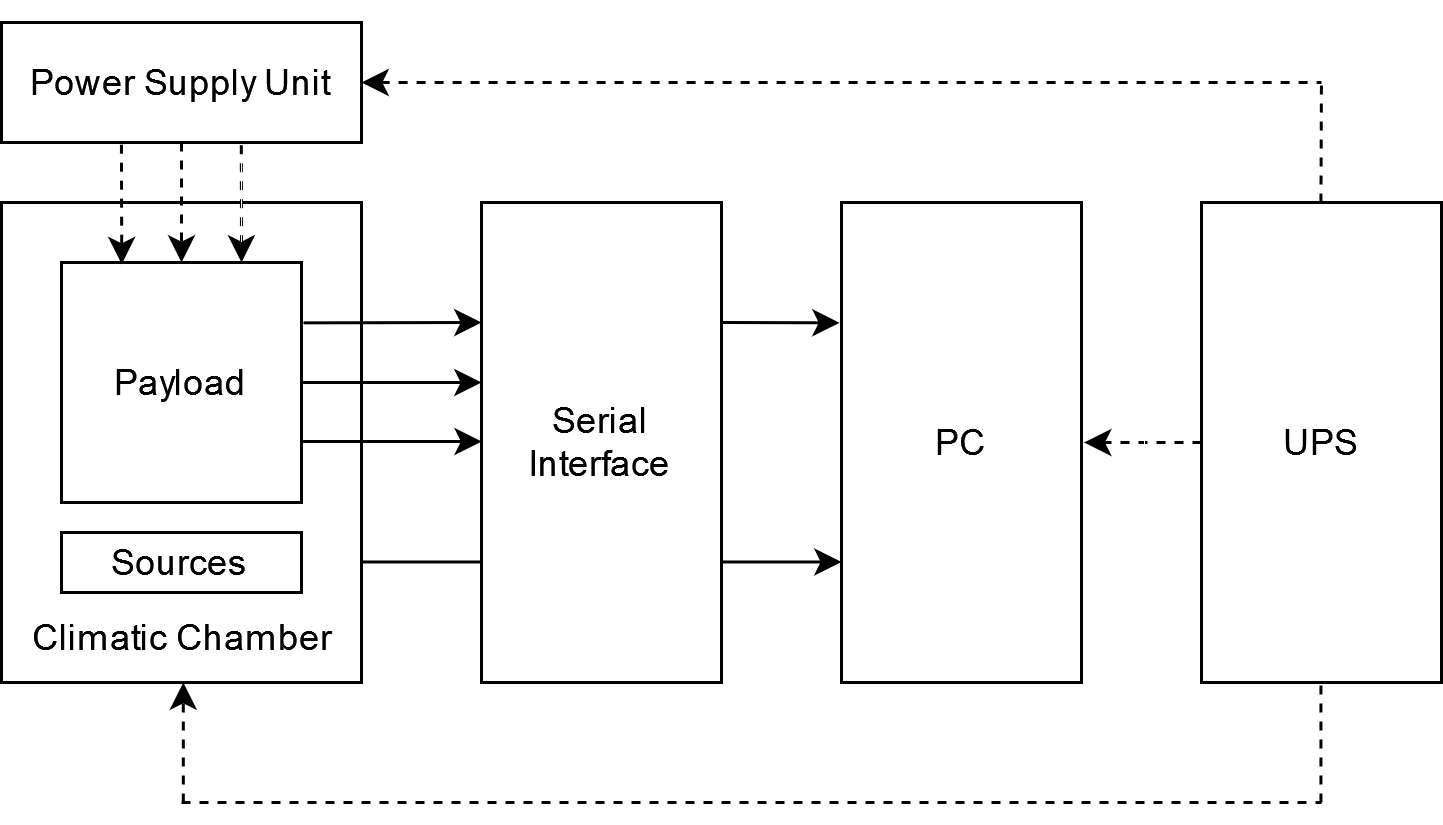}
\caption{Block diagram exemplifying the payload calibration setup. Solid and dashed connections represent data communication and power lines, respectively.}
\label{fig:setup_diagram}
\end{figure}

During observation, when a trigger occurs (i.e., when one or more channels detect a photon event with amplitude above a programmable threshold), the BEE commands the digitization of all channels above threshold in the relevant detector quadrant, records the amplitude measured by the ADC, parses a timestamp, and stores all the data in a buffer. The buffers are organized as a list of events and transferred to the PDHU at regular intervals (nominally set to $1$~s), where they are aggregated with housekeeping data from multiple sensors.
The amplitude information in the PDHU data products is expressed in analog-to-digital units (ADU), which have no physical meaning. Through calibration, a number of parameters characterizing the detector are measured. These parameters allow tagging each photon event with a proper energy measurement. To determine the value of these parameters, the payload is exposed to multiple radioactive sources. The recorded events are analyzed and, when appropriate, assigned to a specific decay of the calibration sources, whose characteristic energies are known with high precision.

\begin{figure}
\centering
    \begin{subfigure}{1.\columnwidth}
        \includegraphics[width=1.\textwidth]{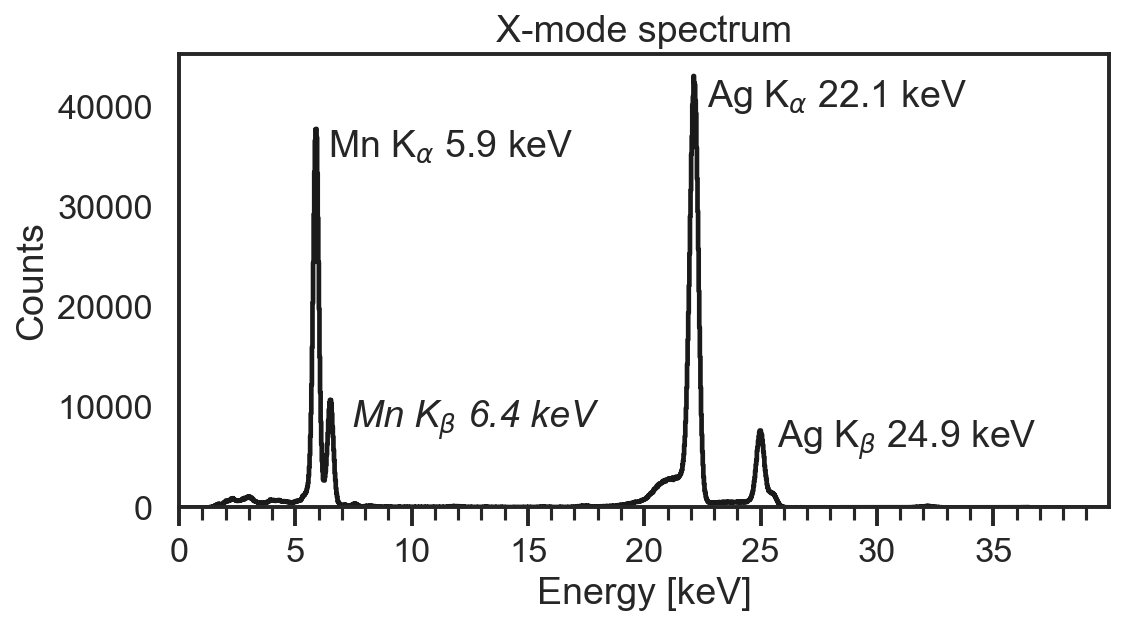}
        \label{fig:dasd}
    \end{subfigure}
    \begin{subfigure}{1.\columnwidth}
        \includegraphics[width=1.\textwidth]{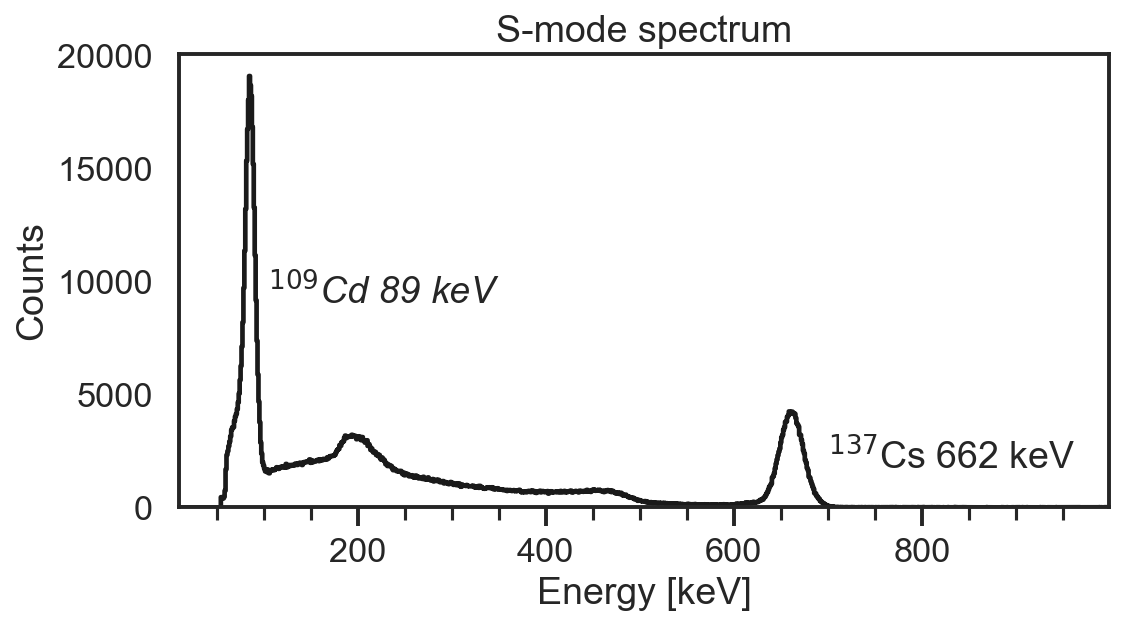}
        \label{fig:asd}
    \end{subfigure}
\caption{Example of a calibration spectrum. The top panel displays the X-mode instrument spectrum, while the bottom panel displays the S-mode. The calibration sources used were $^{55}$Fe, $^{109}$Cd, and $^{137}$Cs. This acquisition was obtained using payload FM3 at a temperature of 0~$^\circ$C.}
\label{fig:example-spectra}
\end{figure}

The calibration campaign took place at INAF-IAPS, through 2022 and early 2023. The experimental setup is represented schematically in \fig\ref{fig:setup_diagram}. The payload was placed in a climatic chamber and exposed to the radioactive sources $^{55}$Fe,  $^{109}$Cd, and $^{137}$Cs. The emission lines produced by these sources and used for the calibration are the Mn K$_\alpha$ (5.9 keV), the Ag K$_\alpha$ (22.1 keV), the Ag K$_\beta$ (24.9 keV) and the $^{137}$Cs 662~keV. The payload was connected to a desktop computer through a custom-made interface that redirected the UART, RS422, and SPI communications to a USB link. Payload operations (e.g., operative mode transitions, data transfer, and file system management) were commanded through a software developed specifically for this purpose. The payload power was supplied by a laboratory power supply unit over three channels with voltages of $3.3$~V, $5$~V and $12$~V. The power supply, the climatic chamber and the desktop computer were powered through an uninterruptible power supply and the entire experimental setup was housed in an ISO-7 (FS-209 class 10000) clean room.
Data were acquired for all payloads, at temperatures of $-$20~$^\circ$C, $-$10~$^\circ$C, 0$^\circ$C, +10~$^\circ$C, +20~$^\circ$C and for three different energy threshold settings. These data were parsed into tables and analyzed by a software pipeline implementing the numerical recipe for calibration \cite{DILILLO2024100797}. 

The calibration recipe is the following:
\begin{enumerate}
    \item {
    \emph{Event tagging and filtering.}
    The observed events are classified as either ``X" or ``S" based on the detection mode. The ``X" tag is used to represent low-energy active SDD detections, while the ``S" tag is used for scintillator events. Events defying this classification (i.e., events comprising triggers over more than two channels, or over two uncoupled channels) are filtered out of the dataset.
    }
    \item {
    \emph{SDD calibration.}
    The X emission lines (Mn K$_\alpha$ 5.9 keV, Ag K$_\alpha$ 22.1 keV and Ag K$_\beta$ 24.9 keV) are located in the X-tagged spectra, separately for each channel. A linear regression is performed between the detected X emission line centroids $A_i$ and the emission line energies $E_i$:
    \begin{equation}
    A_i [\mathrm{ADU}] = \mathrm{Gain} \cdot E_i [\mathrm{keV}] + \mathrm{Offset}
    \end{equation}
    The regression free parameters are called \emph{gain} (slope) and \emph{offset}. The \emph{gain} and \emph{offset} characterize the response of the X spectroscopic chain. This step allows for the amplitude of the X-tagged events to be expressed in energy units, thereby completing their calibration process.
    }
    \item {
    \emph{Photoelectron conversion.}
    Assuming that one electron-hole pair is released into the silicon for each $3.65$ eV of incident radiation energy~\citep{mazziotta2008electron}, the measurement of the gain and offset parameters enables expressing the events amplitude in units of photoelectrons. The amplitude of the S-tagged events are converted to electrons, using the gain and offset parameters estimated from the X calibration of the appropriate channel, following:
    \begin{equation}
        A [\mathrm{e}^{-}] = \frac{10^{3}}{3.65} \frac{A [\mathrm{ADU}] - \mathrm{Offset}}{\mathrm{Gain}}
    \end{equation} 
    }
    \item {
    \emph{Scintillator calibration.}
    This last step is leveraged to gauge the total number of photoelectrons produced in an SDD pair by each scintillation event. 
    Coincident, S-tagged events from coupled channels are equalized, aggregated and summed. The energy peaks due to the $\gamma$ emission lines ($^{137}$Cs 662~keV) are located in each scintillator spectra and the scintillator effective light output is computed according to:
    \begin{equation}
    \label{scalib1}
    LY[\mathrm{e}^-/\mathrm{keV}]=\frac{A [\mathrm{e}^{-}]}{A_{662}[\mathrm{keV}]}
    \end{equation}
    where $LY$ is the effective light output for the scintillator and $A_{662}=661.67$~keV. 
    }
    Finally, an \emph{effective light output} value is assigned to each channel through:
    \begin{equation}
    \frac{LY_1}{LY} = \frac{A_1}{A}
    \quad\mathrm{and}\quad 
    \frac{LY_2}{LY} = \frac{A_2}{A}
    \end{equation}
    where $LY$ is the scintillator effective light-yield, $A_1$ and $A_2$ are the centroids of the calibration line in the individual channels spectra, and $LY_1$ and $LY_2$ are the effective light output measured for the channel couple. This operation concludes the calibration of the ``S"-tagged events.
\end{enumerate}
A typical calibrated spectrum is displayed in \fig\ref{fig:example-spectra}.

\section{Results}
\label{sec:results}
The HERMES payloads are expected to operate between $-$20~$^\circ$C and +20~$^\circ$C, with most of the operational time spent near 0~$^\circ$C.
Unless indicated otherwise, the results of this section refer to data gathered at a temperature of 0~$^\circ$C.

\begin{table*}
\centering
\begin{tabular}{lrrrrrrrrr}
\toprule
 & & & & & \multicolumn{3}{c}{Energy Resolution [eV]} & \multicolumn{1}{c}{[\%]} \\ \cmidrule(r){6-8} \cmidrule(r){9-9}
 & $f_X$ [\%] & $f_S$ [\%] & $T_X$~[keV] & $T_S$~[keV] & $5.9$~keV & $22.1$~keV & $24.9$~keV & $662$~keV \\
\midrule
PFM & 91.7 & 83.3 & 1.7 & 34.9 & 374 $\pm$ 3 & 475 $\pm$ 3 & 502 $\pm$  5 & 4.79 $\pm$  0.02 \\
FM1 & 82.2 & 68.9 & 2.1 & 45.0 & 390 $\pm$ 4 & 483 $\pm$ 4 & 508 $\pm$  7 & 4.86 $\pm$  0.02 \\
FM2 & 83.3 & 70.0 & 2.1 & 43.2 & 317 $\pm$ 5 & 426 $\pm$ 3 & 434 $\pm$  6 & 4.73 $\pm$  0.02 \\
FM3 & 90.0 & 80.0 & 2.2 & 42.5 & 300 $\pm$ 3 & 419 $\pm$ 3 & 438 $\pm$  5 & 4.74 $\pm$  0.03 \\
FM4 & 90.0 & 80.0 & 2.6 & 46.2 & 295 $\pm$ 3 & 414 $\pm$ 3 & 435 $\pm$  6 & 5.11 $\pm$  0.04 \\
FM5 & 90.0 & 80.0 & 2.2 & 43.5 & 295 $\pm$ 3 & 413 $\pm$ 3 & 432 $\pm$  6 & 4.82 $\pm$  0.03 \\
FM6 & 93.3 & 83.3 & 2.0 & 39.1 & 293 $\pm$ 3 & 413 $\pm$ 2 & 423 $\pm$  5 & 4.74 $\pm$  0.02 \\
\bottomrule
\end{tabular}
\bigskip
\caption{A summary table of the payload performances measured at 0~$^\circ$C, reporting the calibrated channel fraction in X- ($f_X$) and S-mode ($f_S$), the X-mode energy thresholds ($T_X$), the S-mode energy thresholds ($T_S$), and FWHM energy resolution as estimated from the emission lines Mn K$_\alpha$ 5.9 keV, Ag K$_\alpha$ 22.1 keV, Ag K$_\beta$ 24.9 keV and $^{137}$Cs 662~keV.} \label{tab:long_tab}
\end{table*}

\paragraph{Calibrated channel fraction} 
In \tab\ref{tab:long_tab} we report the fraction of correctly calibrated channels in both X-mode and S-mode. In X-mode, for all payloads except FM1, the fraction of calibrated channels spans from $82.2 \%$ to $93.3 \%$, which corresponds to a surface ranging from $45.0$~cm$^2$ to $50.4$~cm$^2$, out of the total $54.015$~cm$^2$ detector geometric area. On the other hand, the S-mode calibrated fractions range from $70.0 \%$  to $83.3 \%$. Due to a hardware failure, the FM1 payload was delivered for integration on the SpIRIT satellite with only three active instrument quadrants out of four, resulting in a smaller geometric area of $40$~cm$^2$. Of this surface, $82.2$\% and $68.9$\% were calibrated in X- and S-mode, respectively.
The S-mode calibration requires the X-mode calibration to have been completed for both the channels reading out the crystal. For this reason, the S-mode calibrated fractions cannot be greater than their X-mode counterparts. Most uncalibrated channels were too noisy to be operated at standard threshold levels, and were turned off during data acquisition. This noise is primarily caused by afterpulses induced by the reset signal used to discharge the preamplifier feedback capacitance after a trigger event, and manifests as short-lived bursts in background counts.

\begin{figure}
\centering
        \includegraphics[width=1\columnwidth]{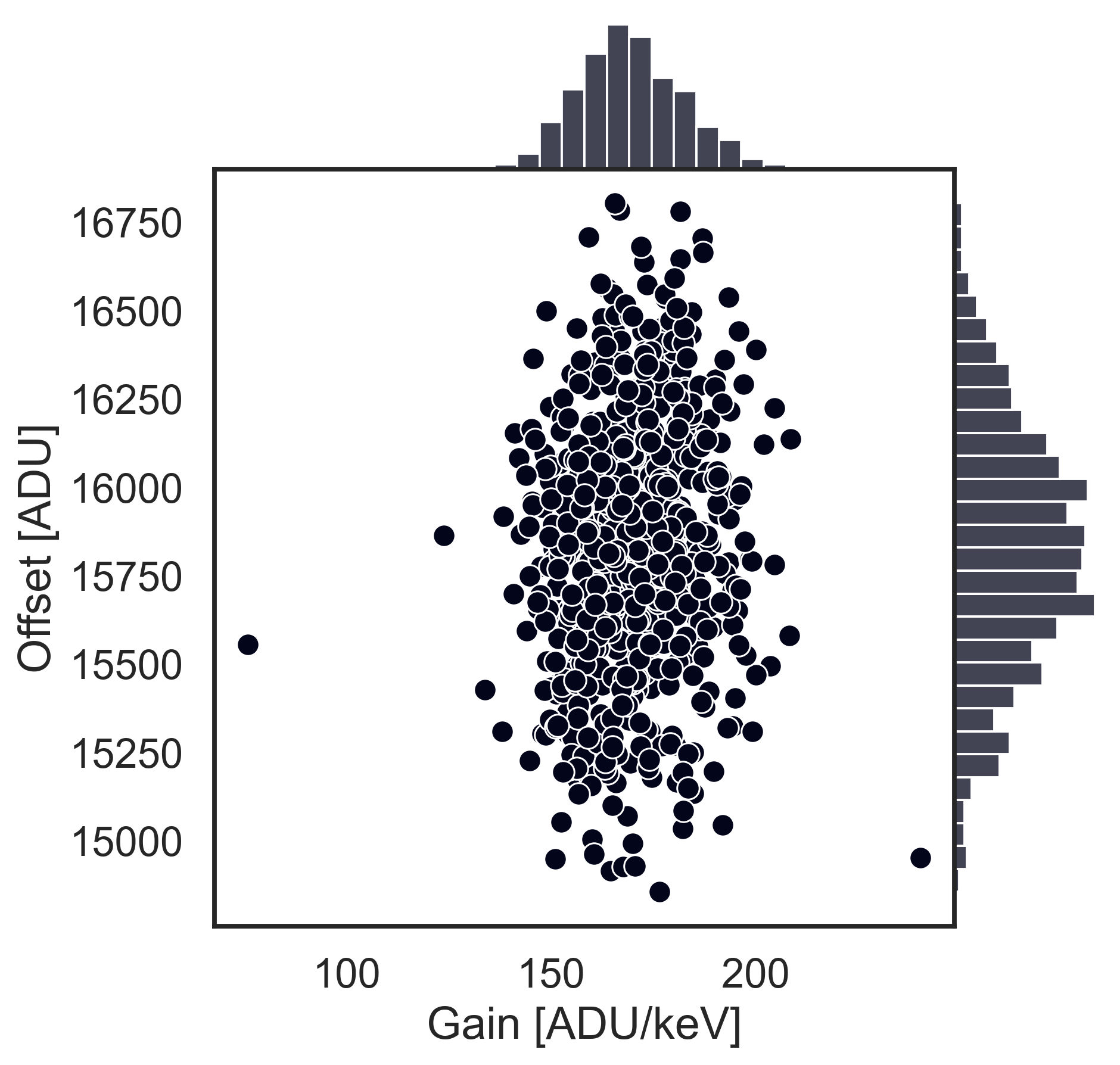}
        \caption{All channel gains and offsets (points) and their marginal distributions (histograms), for all payloads, at a temperature of 0~$^\circ$C.  }
        \label{fig:dispersion}
\end{figure}

\paragraph{Gain, offset, linearity}

\begin{figure*}
\centering
    \begin{subfigure}{.32\textwidth}
        \includegraphics[width=\textwidth]{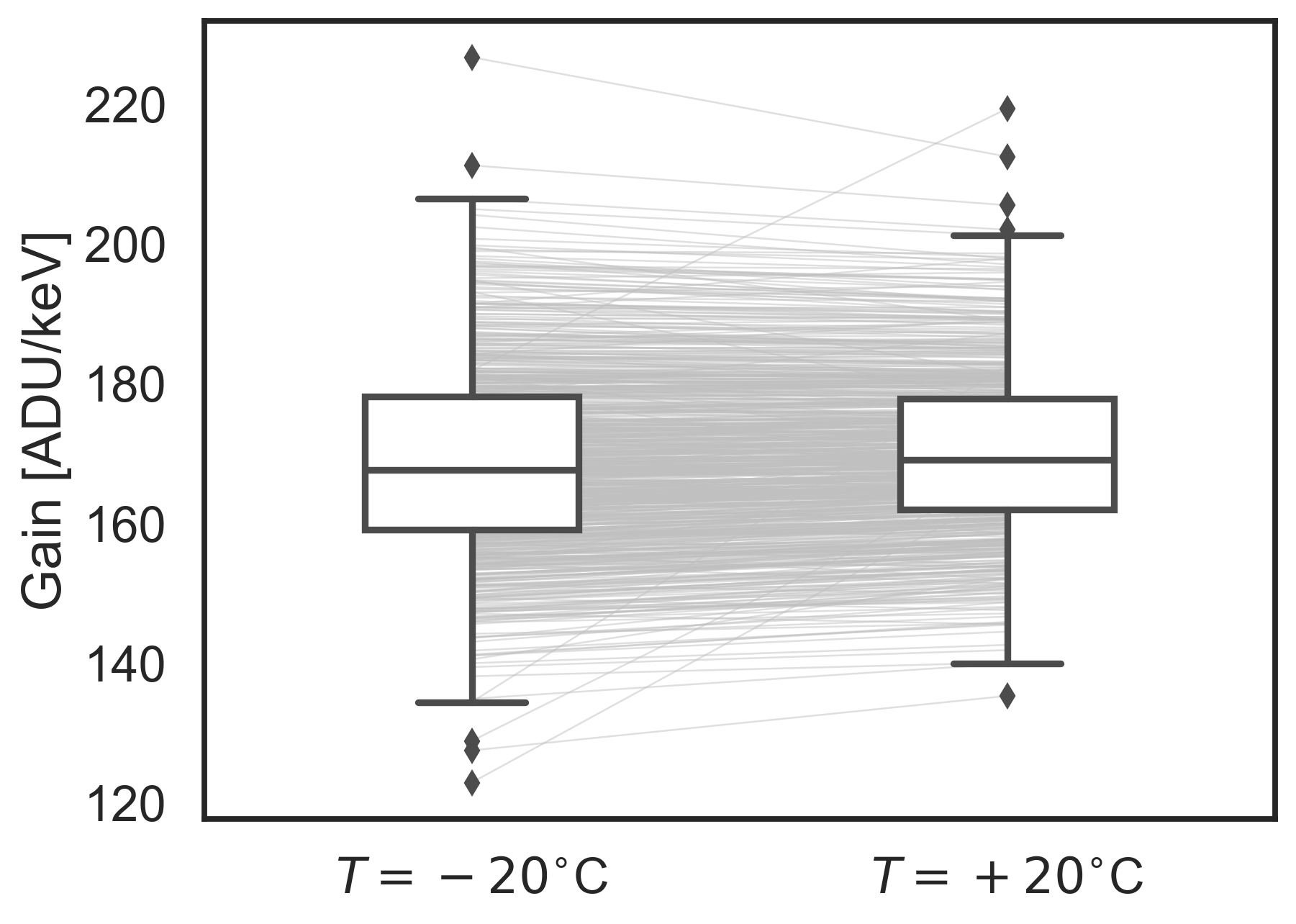}
    \end{subfigure}
    \begin{subfigure}{.34\textwidth}
        \includegraphics[width=\textwidth]{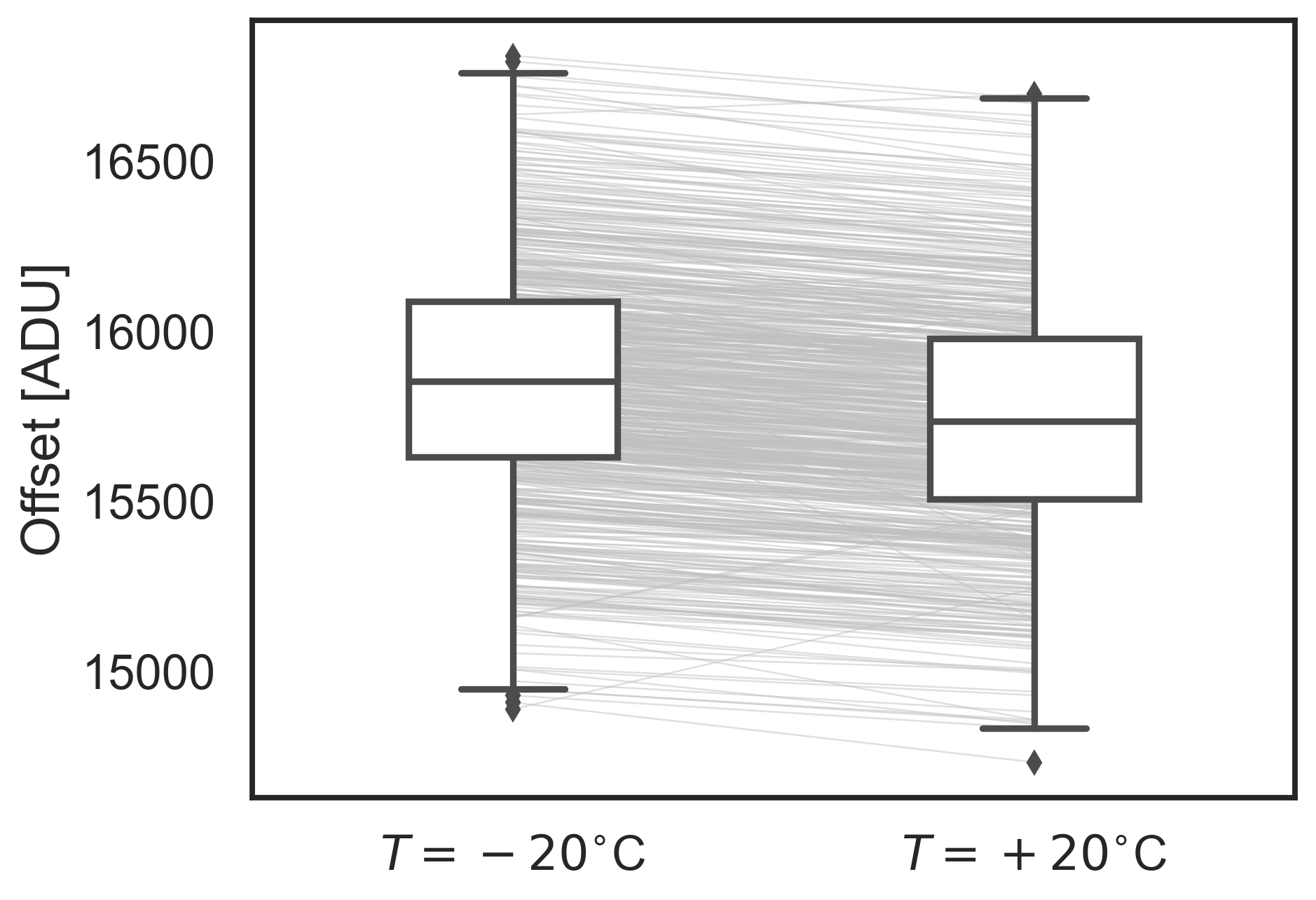}
    \end{subfigure}
    \begin{subfigure}{.32\textwidth}
        \includegraphics[width=\textwidth]{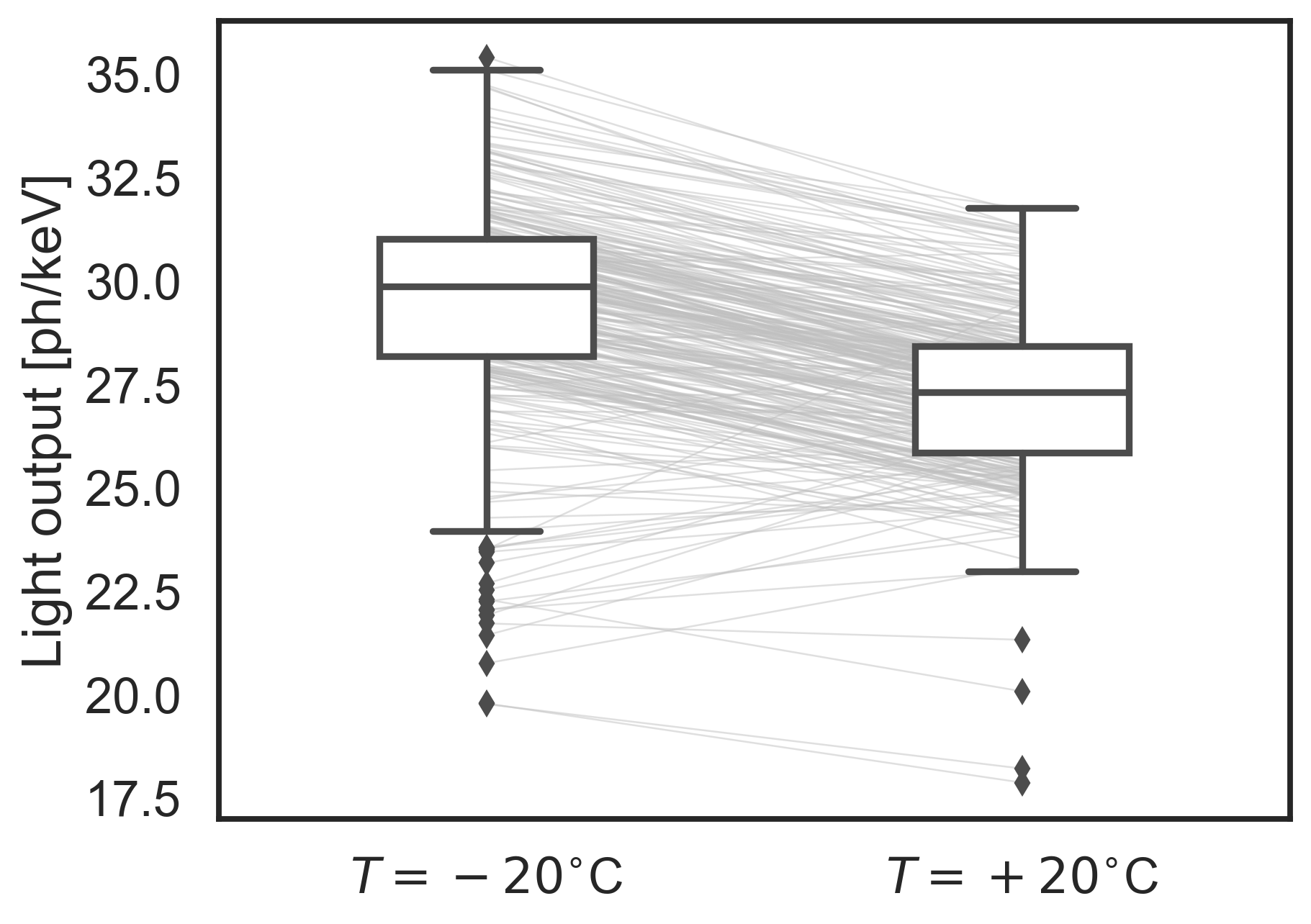}
    \end{subfigure}
\caption{Distributions of gain, offset and light output calibration parameters at +20~$^{\circ}$C and $-$20~$^{\circ}$C, across all payloads. Gray lines are used to connect values for the same channels. }
\label{fig:tempdep}
\end{figure*}

The gain and offset parameters are determined by the characteristics of the spectroscopic chain components. Hence, some level of dispersion is expected due to the finite tolerance of the electronic components and variations in manufacturing processes. 
The parameters measured for each of the 720 calibrated channels from all payloads are shown in \fig\ref{fig:dispersion}. The marginalized gain and offset distributions have mean equal $169.5$ ADU/keV and $15824$ ADU, respectively, and standard deviations $13.6$ ADU/keV and $356$ ADU. A weak positive correlation between gain and offset is observed, with a Pearson correlation coefficient of $r_{xy} = 0.12$ (p-value $=0.002$). 
For 95\% of the channels calibrated in X-mode, the difference between the calibrated peak centers and the true decay energy is smaller than $0.72\%$, compatibly with the LYRA ASIC linearity error requirements and simulations~\cite{gandola2019lyra}.
\fig\ref{fig:tempdep} shows the paired distributions of the calibration parameters at $-$20~$^\circ$C and +20~$^\circ$C for $715$ channels that underwent a successful calibration at both temperatures.
The variation of the median channel gain between  +20~$^\circ$C and  $-$20~$^\circ$C is small, approximately $1\%$. 
For the offsets, we observe a median difference of $109$ ADU over the same temperature range. 

\paragraph{Scintillator light output}
The 312 scintillators for which the S-mode calibration was completed feature a median light output of $29.1$~photons$/$keV with a standard deviation of $2.3$~photons$/$keV. The distribution of the light output measured at temperatures of $-$20~$^\circ$C and +20~$^\circ$C are displayed in \fig\ref{fig:tempdep}.  For most scintillators, the light output increases at low temperatures. This is expected from the physics of the GAGG:Ce scintillator material, and variations up to about $20\%$ in this temperature range have been described in the literature for single-sample studies~\cite{yoneyama2018evaluation}. For our sample, we observe a median difference in light output of $2.5$ photons$/$keV between temperatures $-$20~$^\circ$C and +20~$^\circ$C, corresponding to a $9.4\%$ percentage improvement. 
We note that a secondary effect could contribute to an apparent increase in the measured light output at low temperatures. As temperature decreases, the mobility of electrons inside the SDD silicon bulk increases. In turns, this results in a decrease of the detector ballistic deficit. A smaller ballistic deficit may result in more photoelectrons from scintillation being actually collected by the electronic chain, hence an apparent increase of the scintillator light output. 
For $26$ scintillators, the light output decreases at lower temperatures, which, given the present discussion, is counter-intuitive. Our best explanation for this effect is of mechanical nature. Upon further inspection, scintillators with decreasing light output at lower temperatures also have the largest observed discrepancies between the individual channel effective light output: one SDD channel seems to collect much more scintillation light than its companion.  This could be due to an uneven optical coupling, which is achieved by compression onto the SDDs through a $\sim$3~mm thick transparent silicone pad. At lower temperatures, thermoelastic deformations of the detector assembly could lead to a worsening of the optical coupling effectiveness, thus causing an apparent reduction in crystal  light output. 

\paragraph{Energy thresholds}

\begin{figure}
\centering
    \includegraphics[width=0.98\columnwidth]{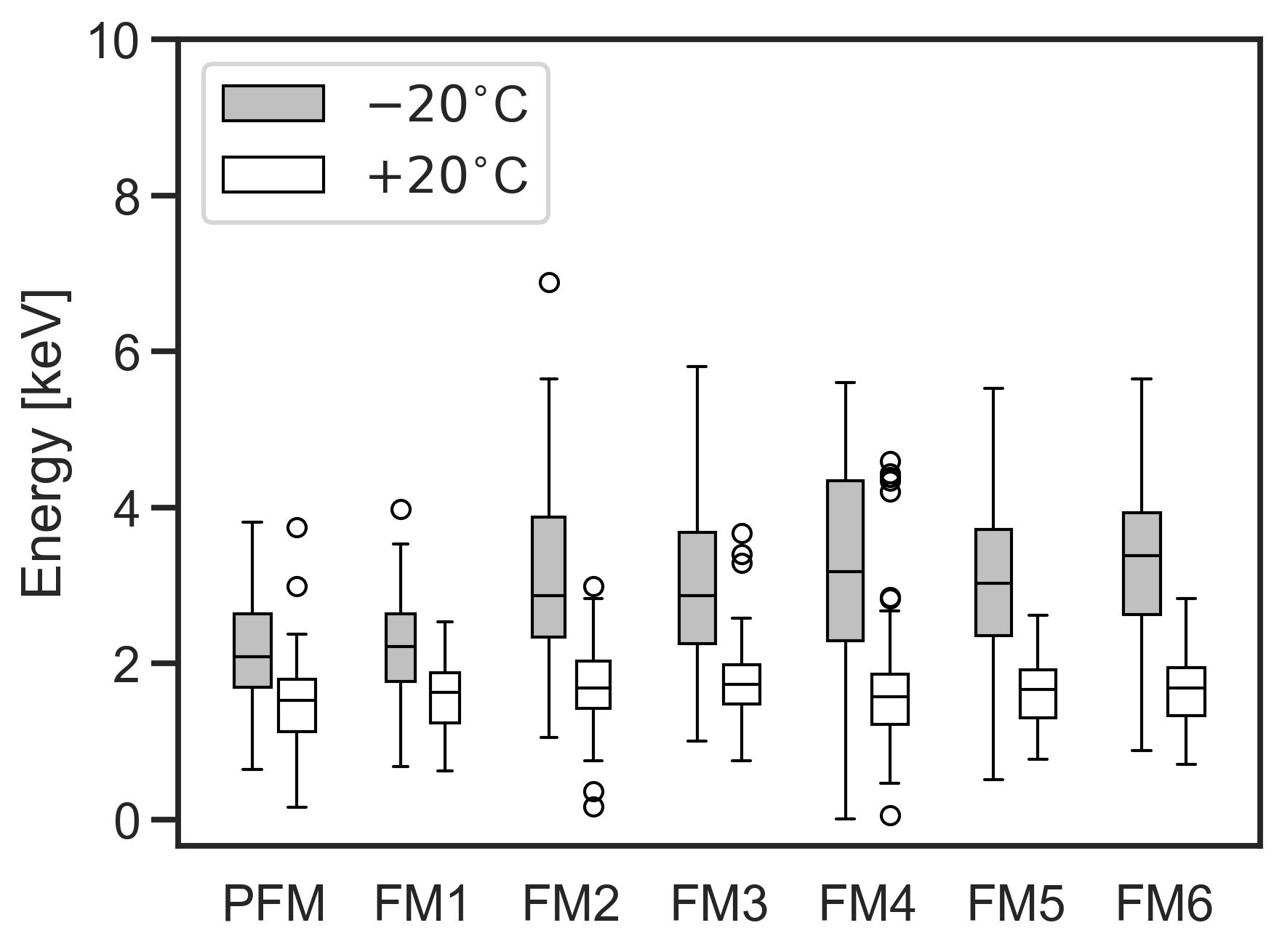}
    \caption{X-mode low-energy threshold at temperature $-$20~$^\circ$C and 20~$^\circ$C, for fixed global DAC threshold. The boxplot width is proportional to the number of correctly calibrated channels from each payload.}
    \label{fig:xthr}
\end{figure}

\begin{figure}
\centering
    \includegraphics[width=\columnwidth]{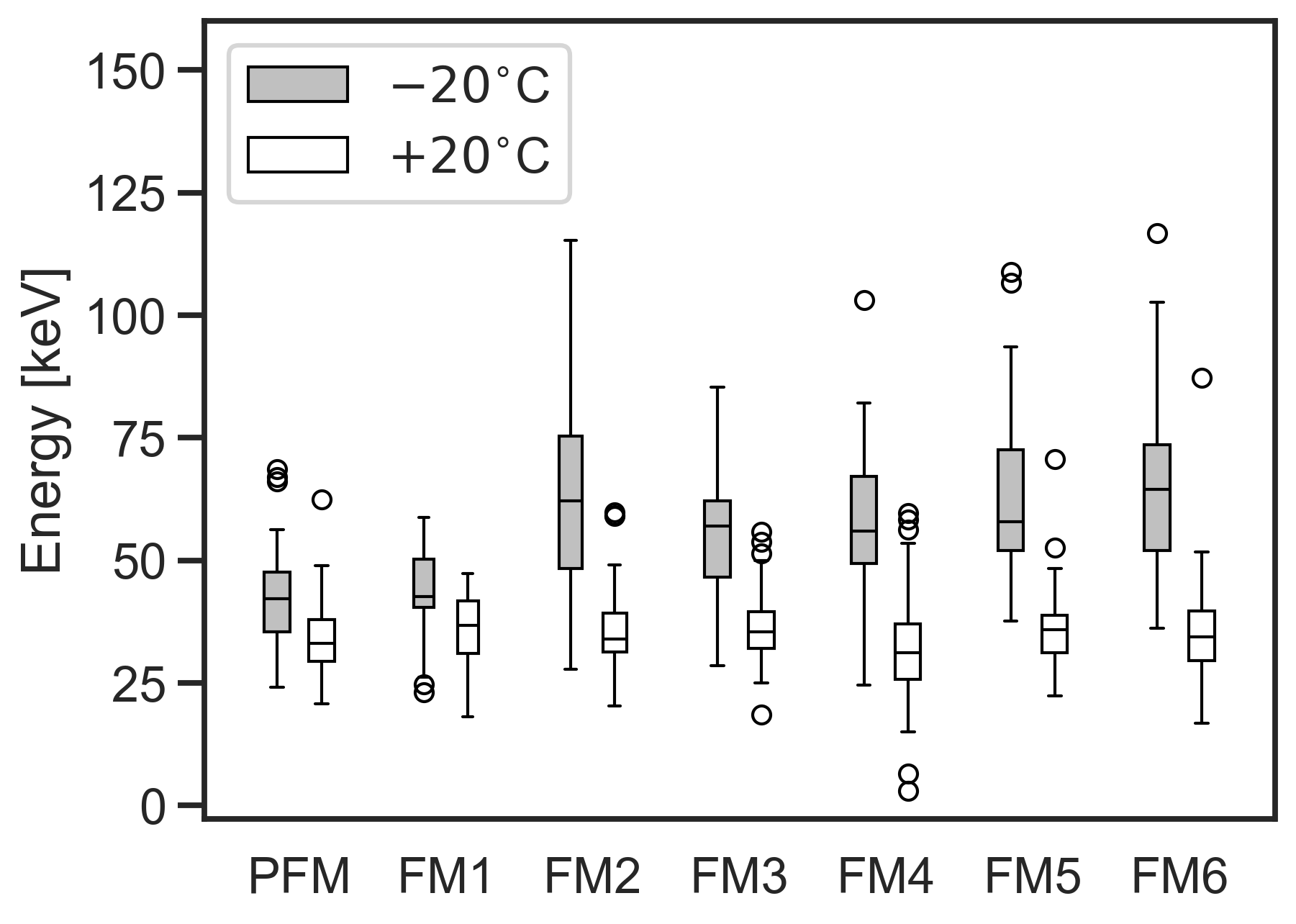}
    \caption{
    S-mode low-energy threshold at temperature $-$20~$^\circ$C and 20~$^\circ$C, for fixed global DAC threshold. The boxplot width is proportional to the number of correctly calibrated channels from each payload.
    }
    \label{fig:sthr}
\end{figure}

\begin{figure}
\centering
    \includegraphics[width=\columnwidth]{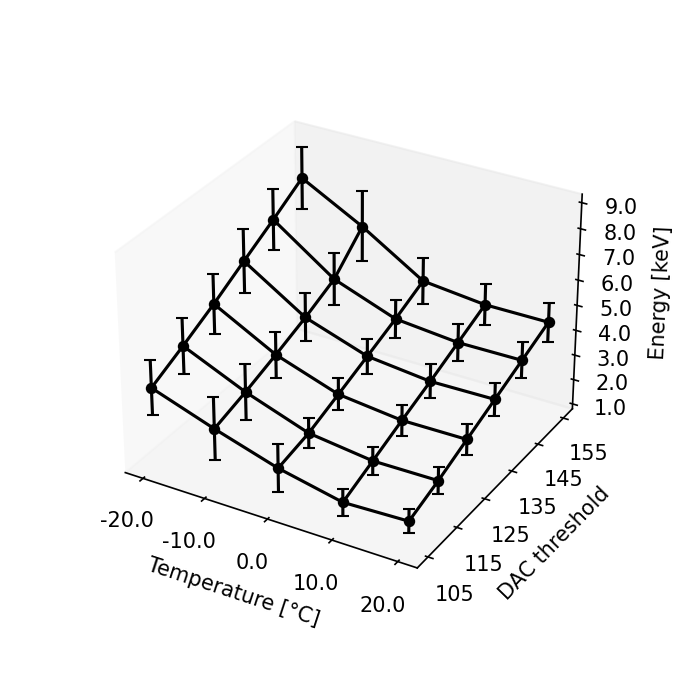}
    \caption{X-mode energy threshold versus temperature and global DAC threshold, as measured for payload FM6. For a fixed DAC threshold, the energy threshold increases at lower temperatures. Lower energy thresholds can be achieved by setting a lower DAC threshold parameter. Error bars represent the standard deviations computed across all calibrated channels.}
    \label{fig:thrsurface}
\end{figure}

The detector energy threshold represents the minimum energy a photon must have to be detected, either in X- or S-mode. It is ultimately determined by the electronics trigger threshold setting and the specific characteristics of each channel, such as gain, offset, and light output.
The trigger threshold is configured using a LYRA-BE ``global" DAC setting, which is common to all LYRA-BE channels, along with ``fine" threshold settings that enable further tuning of individual channels. 
The combined global and individual thresholds are set sufficiently high to effectively suppress noise-induced triggers in most detector channels.
The median values of the channel threshold distributions represent the minimum energy at which half of the detector is operational and are summarized in \tab\ref{tab:long_tab}.
As shown in \fig\ref{fig:xthr} and \fig\ref{fig:sthr}, we observe an increase in the energy threshold at constant DAC threshold setting for lower temperatures. This dependence has been verified for all the HERMES payloads and applies to both the X-mode and S-mode events.
Being the parallel noise (induced by leakage current) smaller at lower temperatures and the GAGG crystal light output higher, in order to exploit the better system noise performance at low temperatures, we performed a specific calibration campaign aimed at estimating the temperature-dependent variation of the DAC threshold and thus providing the best threshold setting at different operating temperatures.
Indeed, by appropriately adjusting this parameter, it is possible to achieve lower energy thresholds compared to those attainable at higher temperatures, as demonstrated for X-mode events in \fig\ref{fig:thrsurface}.

\begin{figure*}[h]
\centering
    \includegraphics[width=0.97\textwidth]{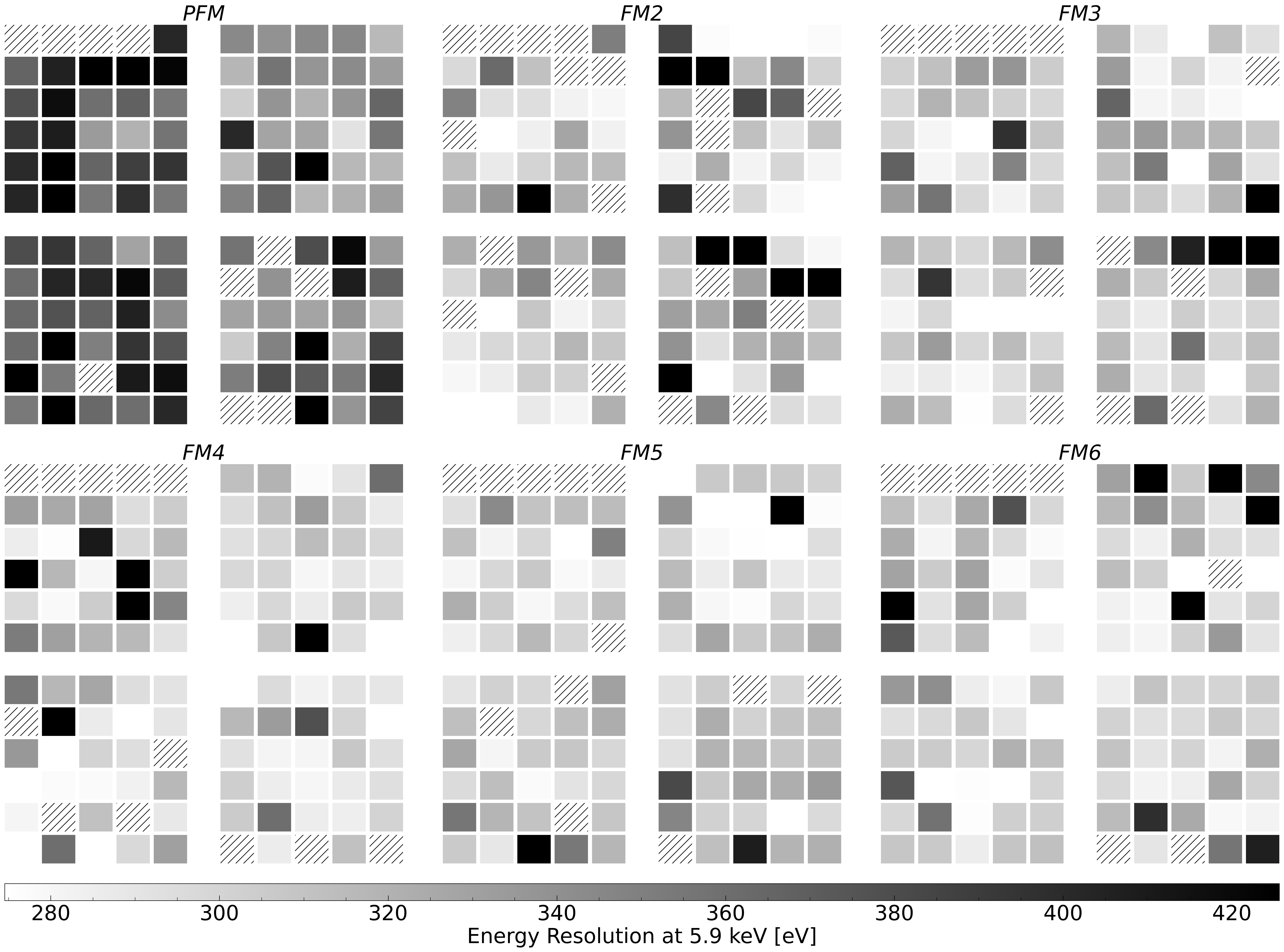}
    \caption{FWHM energy resolution at $5.9$~keV and a temperature of 0~$^\circ$C. Values from different channels are mapped to the channels position on the HERMES-Pathfinder detector planes.}
    \label{fig:resmap_nofm1}
\end{figure*}
\begin{figure}[h]
\centering
    \includegraphics[width=.88\columnwidth]{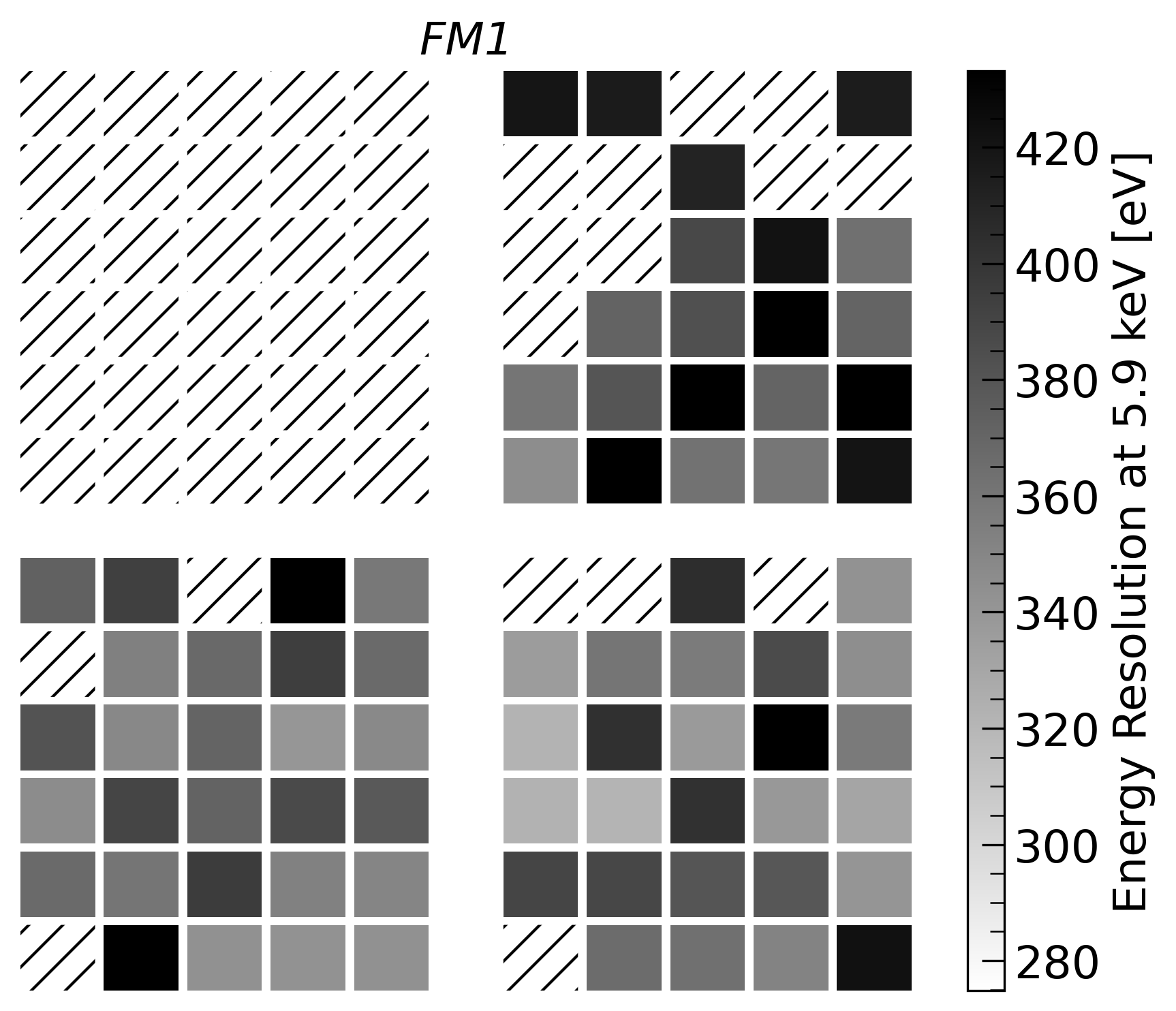}
    \caption{As with \fig\ref{fig:resmap_nofm1}, FM1-SpIRIT payload.}
    \label{fig:resmap_fm1}
\end{figure}

\paragraph{Spectroscopic resolution}
The X-mode resolutions measured over the integrated calibration spectra are summarized in \tab\ref{tab:long_tab} and span from $293$ eV to $390$ eV FWHM for the Mn K$_\alpha$ $5.9$ keV calibration line.  In \tab\ref{tab:long_tab} we also report the spectroscopic performance of the detector in S-mode, as computed over the summed spectra for the $^{137}$Cs 662~keV calibration line. The S-mode resolutions are below $5\%$ for all payloads except for FM4 ($5.11\% \pm 0.04$, under the same conditions). 
In \fig\ref{fig:resmap_nofm1} and \fig\ref{fig:resmap_fm1} the FWHM energy resolution computed at Mn K$_\alpha$ $5.9$ keV is shown for all the X-mode calibrated channels, mapping the resolution to the channel position on the detector planes, for all HERMES payloads.
No systematic patterns are observed when examining the resolution of channels occupying identical positions on the detector planes of different payloads, except for the PFM and FM1 payload, which exhibit inferior spectroscopic resolution compared to other flight models.

\paragraph{Counting}

\begin{figure}
\centering
    \includegraphics[width=\columnwidth]{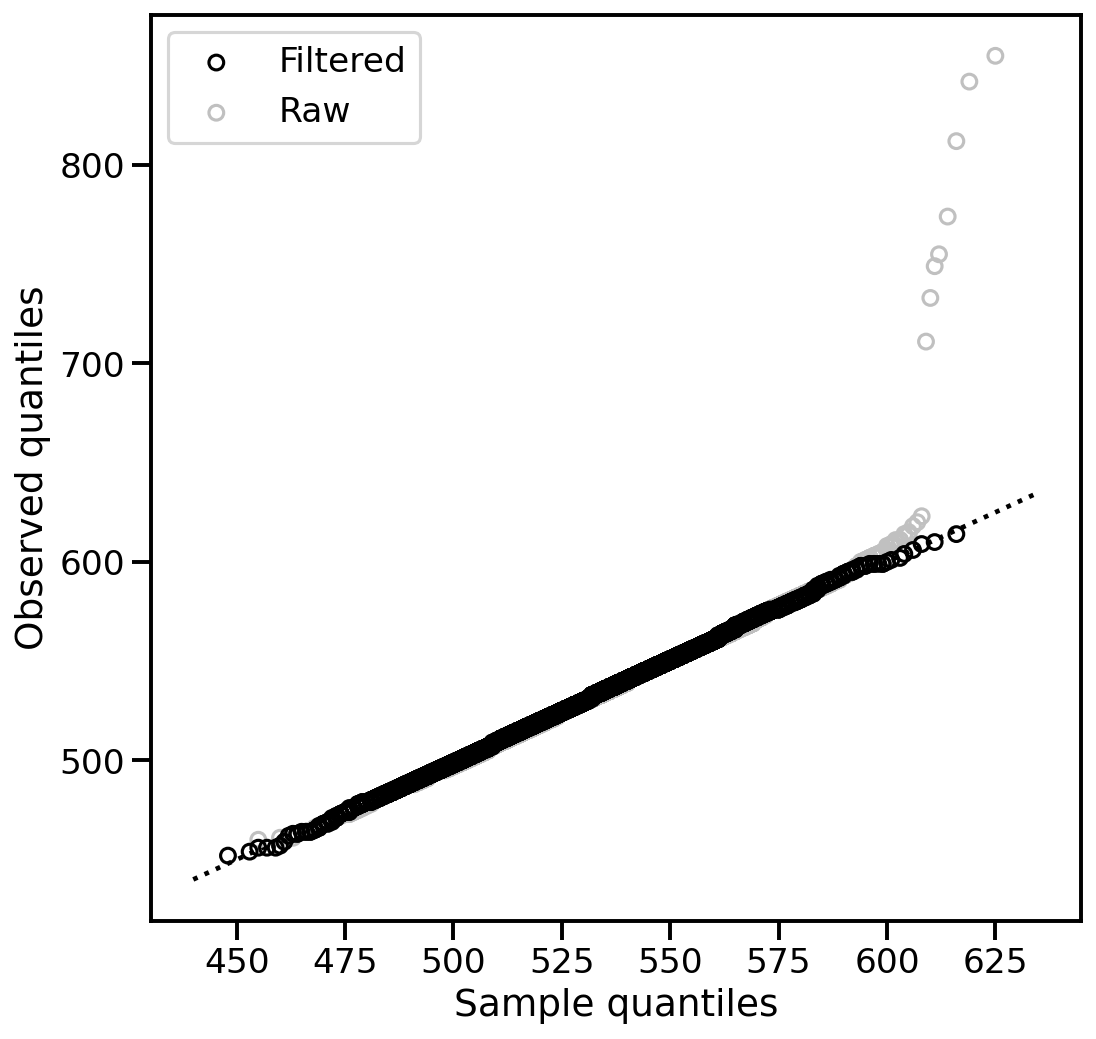}
    \caption{Quantile-quantile plot of the detector counts observed through a $600$ s calibration acquisition of a radioactive source with constant intensity, versus an equal number counts sampled from a Poisson distribution with the same mean rate. Data are binned with $0.1$ s bin-length.}
    \label{fig:qqplot}
\end{figure}

Quantile-quantile plots are graphical tools useful for determining if two data sets come from populations with a common distribution. A point in a quantile-quantile plot corresponds to one of the quantiles of a data set, plotted against the same quantile of the other data set. Presently, we apply this technique to detect deviations from the Poisson distribution in the detector lightcurves. Such deviations may result from instrumental effects undermining the basic assumptions of a Poisson model, most importantly the independence of different events.
\fig\ref{fig:qqplot} shows a quantile-quantile plot of the detector counts observed during one calibration acquisition versus an equal number of counts sampled from a Poisson distribution with the same mean rate.
The statistic of the observed counts and the control sample agrees over most of the count range, with the exception of a few, high-rate events. These events correspond to random, short-lived bursts in background counts and are caused by the afterpulse noise occasionally induced by the signal responsible for resetting a channel after a trigger. Since the delay $t_0$ between the trigger and the reset signal is known, these spurious bursts are straightforward to identify and filter: it is enough to cut-off events whose delay from the precedent one is equal to or shorter than $t_0$.
This technique could potentially result in censoring source events with small delays. However, since $t_0$ is almost equal to the detector dead time (which ranges from $14.2$~$\mu$s to $22.8$~$\mu$s, depending on the number of simultaneously triggered channels), most of these events would be lost regardless. Filtering will not be performed in real-time, during the satellite operations.  Nevertheless, discrimination between spurious noise bursts and real bright transients can be achieved by other means, such as a coincidence gate comparing count rates from different, independent detector quadrants, a technique commonly employed by spaceborne GRB burst monitors \cite{dilillo2023gamma}. 
Once spurious afterpulse bursts in noise are addressed, counts from the HERMES payload fit well with Poisson statistics, with no signs of over- or under-dispersion, at least up to the highest count rates tested through calibration, which amounts to few thousand counts per second, a value comparable to that expected from a extremely bright GRB. 

\subsection{Discussion}

The HERMES detector was designed to achieve X-mode energy thresholds in the range of a few keVs. This is lower than the typical energy threshold of space-borne GRB monitor experiments. For example, the Fermi-GBM NaI detectors cover an energy range of 8 keV to 1 MeV~\cite{meegan2009fermi}, while Swift-BAT is sensitive in the 15--150 keV range \cite{barthelmy2005burst}. This feature is driven by a scientific motivation. The low-energy end of GRB prompt emission spectra differs from what is expected from the traditional synchrotron model of the GRB radiative mechanism, and alternative models have been proposed to solve this conundrum. These models predict a hardening of the GRB low-energy spectral index in the decade spanning $1$ to $10$ keV, evidence of which has already been found~\cite{oganesyan2017detection}. However, thus far the scope of these studies has been limited by the scarcity of the GRB sample for which spectral information on the prompt emission is available in the soft-X energy band~\cite{oganesyan2018characterization}.
The low X-mode energy threshold achieved by the HERMES detectors has the potential to provide new insights on this little-explored domain of GRB prompt emission. However, there is still margin for improvement. The X-mode energy thresholds are theoretically limited by the electronic noise of the spectroscopic chain. In our case however, the value of the thresholds appears to be dominated by other sources. To elaborate on this point it is worthy to consider a numerical case. The statistical noise contribution at $5.9$ keV  is $\approx 14$ e$^{-}$ RMS, assuming that for each eV of incident radiation $3.65$ electron-hole pairs are freed into silicon, and taking into account the Fano factor $\approx 0.117$~\citep{mazziotta2008electron} in silicon. Subtracting this contribution to the FWHM resolution measured at $5.9$ keV by the PFM payload, we obtain an upper-bound estimate of the equivalent electronic noise of $\approx 41.3$ e$^{-}$ RMS, corresponding to a $5\sigma$ threshold of $\approx 750$ eV, a value much smaller than the actual threshold. This implies that, to achieve a functional configuration for the threshold, a number of noise sources different from electronic noise must be considered. 
Our diagnosis is that the main driver for extra noise are the digital channel multiplexing and the signals to reset channels after a trigger. Improving on these aspects could result in a further suppression of the energy thresholds, as well as other aspects of the detector performance such as the calibrated area fraction and dead time.

\section{Conclusions}\label{sec:conclusion}

In this paper, we:
\begin{enumerate}
    \item Describe our recipe for the calibration of the HERMES siswich detector.
    \item Quantify scintillators light output and the gain-offset parameters characterizing the ASIC-SDDs system.
    \item Quantify the low-energy threshold of the instrument, both in X- and S-mode.
    \item Estimate the instrument spectroscopic resolution at multiple energies spanning from $5.9$ keV to $662$ keV.
    \item Discuss the counting properties of the HERMES detector.
\end{enumerate}
Whenever possible, we discuss the dependence of these features on temperature over a range of $-$20~$^\circ$C to +20~$^\circ$C. 

\backmatter

\paragraph{Funding}
This work has been carried out in the framework of the HERMES-TP and HERMES-SP collaborations. We acknowledge support from the European Union Horizon 2020 Research and Innovation Framework Programme under grant agreement HERMES-Scientific Pathfinder n. 821896 and from ASI-INAF Accordo Attuativo HERMES-Technologic Pathfinder n. 2018-10-H.1-2020. 

\paragraph{Author Contributions} 
All authors whose names appear on this submission made significant contributions, either by participating in the experiment design and data collection, providing software required for data analysis, participating in manuscript writing and revision, or providing resources and information that proved critical to the completion of this work.

\section*{Declarations}

\paragraph{Competing Interests}
The authors declare no competing interests.


\bibliography{sn-bibliography}

\end{document}